\documentclass[runningheads]{llncs}
\usepackage[T1]{fontenc}
\usepackage{graphicx}
\usepackage{comment}
\usepackage{listings} 
\usepackage{orcidlink}
\usepackage{hyperref}
\usepackage{eso-pic}
\usepackage{listings} 
\lstdefinestyle{javaStyle}{
  language=Java,
  basicstyle=\ttfamily\small,
  keywordstyle=\bfseries,
  commentstyle=\itshape\color{gray},
  stringstyle=\color{red},
  showstringspaces=false,
  numbers=left,
  numberstyle=\tiny,
  breaklines=true
}

\begin{document}
\AddToShipoutPicture*{%
  \AtPageUpperLeft{%
    \raisebox{-1.6cm}{%
      \hspace*{0.5\paperwidth}%
      \makebox[0pt]{%
        \begin{minipage}{1\textwidth}
        \centering
        \scriptsize\itshape
        Pre-peer-review version submitted to the Workshop Track of ECSA 2025, Springer LNCS 15982.\\
        Final Version of Record: \url{https://doi.org/10.1007/978-3-032-04403-7_13}
        \end{minipage}
      }%
    }%
  }%
}

\title{Towards Benchmarking Design Pattern Detection Under Obfuscation: Reproducing and Evaluating Attention-Based Detection Method}

\titlerunning{Benchmarking Design Pattern Detection Under Obfuscation}
%
\author{ Manthan Shenoy\inst{1}\orcidlink{0009-0002-2756-1652} \and
Andreas Rausch\inst{1}\orcidlink{0000-0002-6850-6409}}
\authorrunning{M.Shenoy, A.Rausch}
%
\institute{Institute for Software and Systems Engineering,Technische Universität Clausthal, Clausthal-Zellerfeld, Germany \\
\email{\{ms84,andreas.rausch\}@tu-clausthal.de}}
\maketitle              

\begin{abstract}
This paper investigates the semantic robustness of attention-
based classifiers for design pattern detection, particularly focusing on their reliance on structural and behavioral semantics. We reproduce the DPD\textsubscript{Att}, an attention-based design pattern detection approach using  learning based classifiers--and evaluate its performance under obfuscation.
To this end, we curate an obfuscated version of the DPD\textsubscript{Att} Corpus, where the name identifiers in code such as class names, method names, etc and string literals like print statements, comment blocks, etc are replaced while preserving control flow, inheritance and logic. Our findings reveal that these trained classifiers in DPD\textsubscript{Att} depend
significantly on superficial syntactic features, leading to substantial misclassification when such cues are removed through
obfuscation. This work highlights the need for more robust
detection tools capable of capturing deeper semantic meanings in
source code. These findings highlight the need for semantically grounded benchmarks. We propose our curated \textit{Obfuscated corpus} (containing 34 java source files) as a reusable proof-of-concept benchmark for evaluating state-of-the-art design pattern detectors on their true semantic generalization capabilities.


\end{abstract}
\section{Introduction}

Software quality—particularly maintainability and understandability—is a core objective of software engineering. Rana et al. \cite{Rana.et.al.(2021)} performs an empirical analysis demonstrating that the introduction of GoF design patterns yields significant gains in maintainability metrics compared to pattern-free code, while Al-Obeidallah et al.\cite{al2021impact} report that classes embodying design patterns exhibit  maintainability but also higher understandability, thereby reducing code-comprehension effort relative to non-pattern classes.

Despite these benefits, production code rarely contains explicit pattern documentation, forcing developers to spend up to 58\% of their time on program comprehension alone in \cite{xia2017measuring} and contributing to software maintenance effort that can consume up to 70\% of lifecycle costs \cite{o2001inference}. To bridge this gap between pattern theory and practice, the research community has developed several detection techniques that recover architectural intent from legacy systems without relying on explicit documentation.

Early automated detection methods relied on structural matching of code artifacts against rigid templates. Techniques such as similarity scoring between program and pattern graphs in \cite{Tsantalis} and metrics-based detection techniques in \cite{Marinescu} potentially located canonical pattern shapes but proved brittle under minor code variations, yielding high false-positive rates. These structure‐focused methods lacked the capacity to capture the semantic or behavioral nuances that distinguish, for example, two patterns with identical class hierarchies but differing runtime interactions.

 To handle implementation variants as present in real-world systems, reasoning-based frameworks incorporate fuzzy and constraint inference: Niere \cite{Niere} employ fuzzy graph-rewrite rules to allow inexact matches, Rasool et al.\cite{rasool2010design} recover patterns via annotations, regular expressions, and SQL queries over intermediate code representations , and Guéhéneuc et al.\cite{demima} uses explanation-based constraint programming to recognize design-motif variants semi-automatically using their tool called DeMIMA. While these approaches increase recall on variant implementations, they depend on expert-tuned parameters and  manual effort.
 
To reduce manual rule-crafting,  the field shifted toward machine-learning classifiers. Feature-based detectors such as DPD\textsubscript{F} \cite{8feture-based} combine Word2Vec embeddings with ensemble classifiers to achieve over 80\% precision and recall, requiring substantial labeled data.

Recently, attention‑based transformer \cite{vaswani2017attention} architectures were proposed to capture deeper contextual semantics and behavioral logic. Mzid et al. \cite{mzid2024attention} introduce DPD\textsubscript{Att} detector, which first generates contextualized embeddings of code tokens and then applies multi‑head attention to identify semantically salient relationships among classes, methods, and fields. This context‑aware mechanism enables DPD\textsubscript{Att} detector to generalize over real world software systems which they used from \cite{8feture-based} and augmented to having a 1645 Java files spanning 13 GoF patterns , achieving 86\% precision and recall.

In our analysis of the DPDAtt corpus, we identified pervasive \textbf{label leakage}: syntactic cues i.e, pattern names appear directly in class names (e.g., "SingletonManager", "ObserverClient"), as well as in package names, comment blocks, and string literals. Prior work has shown that such syntactic cues can dominate model decisions—Gao et al.\cite{gao2023two} demonstrate that renaming as few as one identifier in source code can cause neural code comprehension models to produce completely irrelevant outputs, despite unchanged program semantics. Similarly, Dunlop et al.\cite{dunlop2023investigating} observe that deep-learning detectors for design patterns misclassify non-pattern code when pattern terms appear in identifiers .

Despite these concerns, current evaluation methods in pattern detection often overlook this weakness. Performance is typically measured using datasets that include these syntactic signals, leaving unanswered whether detectors like DPD\textsubscript{Att} detector truly understand the architectural semantics of design patterns. This emphasized the importance of developing a benchmark that maintains structural and behavioral logic while removing these syntactic signals. An effective benchmark should test if models can generalize across variations in naming and documentation, as commonly found in real-world codebases. 

To address this, We propose a validation framework focused on semantic robustness. Our method involves systematically obfuscating a portion of the DPD\textsubscript{Att}  corpus, eliminating all identifier-based signals while keeping the executable logic intact. This framework offers a structured way to determine if detection models genuinely learn the semantics of design patterns or simply overfit to token-level features. 

To guide our research, we propose the following hypothesis and questions: 

\textbf{Research Hypothesis}: Systematic code obfuscation that removes syntactic signals while preserving structural and behavioral semantics can reveal whether design pattern detectors, such as DPD\textsubscript{Att}, truly capture semantic understanding.

\begin{itemize}
  \item \textbf{RQ1}: How does the performance of machine learning–based detectors such as DPD\textsubscript{Att} degrade when syntactic signals are removed?

  \item \textbf{RQ2}: To what extent can a systematically obfuscated, 13-pattern subset of the {\textit{DPD\textsubscript{Att} Corpus} serve as a valid and reusable benchmark for testing the semantic robustness of design pattern detectors?}
\end{itemize}

The remainder of this paper is organized as follows:
Chapter \ref{chp-2} outlines the methodology, Chapters \ref{chp-3}–\ref{chp-6} describe the experimental setup and results, and Chapter \ref{chp-7} presents the analysis and discussion, followed by conclusion and future work in Chapter \ref{chp-8}.

\section{Methodology}\label{chp-2}
 
\begin{figure}
\centering
\includegraphics[width=1\linewidth, height=8cm]{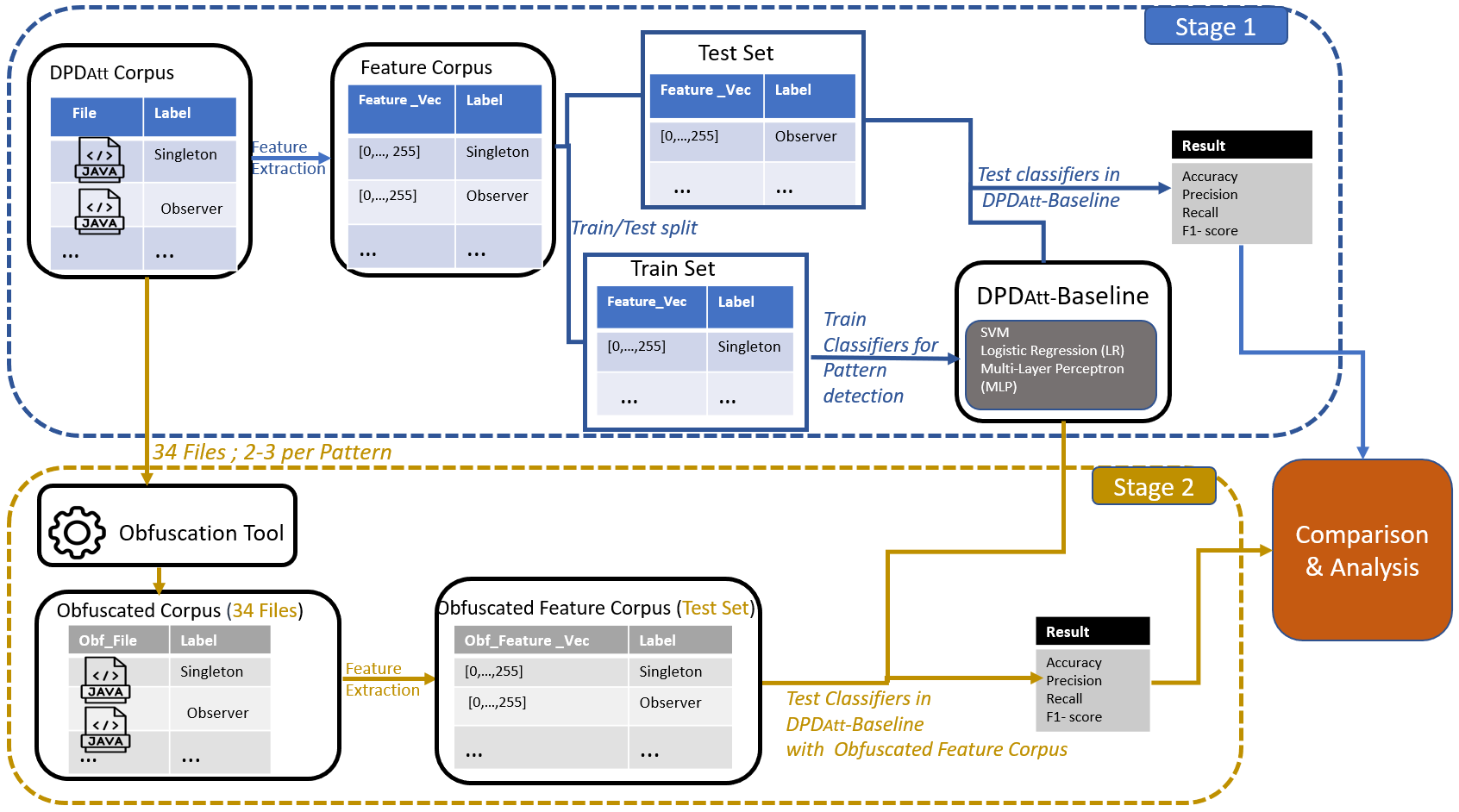}
\caption{Overview of our experimental workflow}
\label{fig:methodology}
\end{figure}

This Chapter presents the methodology adopted in this paper, starting with a recap of the DPD\textsubscript{Att}\cite{mzid2024attention} approach described in \ref{sub-sec2.1} and then providing a high-level overview of the experimental workflow shown in Figure \ref{fig:methodology} to perform comparative analysis of the pre-trained  DPD\textsubscript{Att} classifiers under original and  systematic name-based code obfuscation approach as outlined in \ref{sub-sec2.2}

\subsection{Intro to the DPD\textsubscript{Att} \cite{mzid2024attention}Approach}\label{sub-sec2.1}

The DPD\textsubscript{Att} (Attention method based Design Pattern Detection) \cite{mzid2024attention} approach employs supervised learning classifiers designed to automatically detect the design patterns at the class level. It leverages  attention-based code embedding model to automatically generate the feature vectors from the Java source files. These feature vectors are then used to train different classifiers for pattern classification.

The original DPD\textsubscript{Att} workflow involved:

\begin{itemize}
    \item Curating a large java corpus called \textit{DPD\textsubscript{Att} Corpus}  (containing 1645 Java source files) that covers 13 Gang of Four (GoF) \cite{gamma1995design} design pattern across the Creational (e.g., Singleton, Factory Method,etc), Structural (e.g., Adapter, Decorator,etc), and Behavioral (e.g., Observer, Strategy,etc) categories and "Unknown" (which does not belong to any pattern) for better generalization.
    \item Automatic Transformation of each Java source file into fixed-length feature vector (300-dimensional) using transformer-based code Large language Model(LLM), they specifically use codeT5+ (110M parameter)\cite{wang2023codet5plus} an attention-based encoder that captures statically syntactic and semantic code features. 
    \item Training of multiple classifiers (such as SVM, Logistic Regression, Multi Layer Perceptron) for pattern classification using these feature vectors. 
    \item Evaluating these trained classifiers to compare with state-of-art-approaches using  classification metrics such as precision, recall and F1-score.

\end{itemize}

\subsection{Description of Our experimental workflow}\label{sub-sec2.2}

Figure \ref{fig:methodology} illustrates the complete workflow of our experimental setup, which is structured into \textbf{two main stages}: (1) training and evaluating the baseline design pattern classifiers mentioned in \ref{sub-sec2.1} (we refer as DPD\textsubscript{Att}-Baseline), and (2) evaluating the semantic robustness of these classifiers under systematic code obfuscation.

In \textbf{Stage (1)}, We begin with the reusing the \textit{DPD\textsubscript{Att} Corpus} mentioned in \ref{sub-sec2.1}, is a Java corpus consisting of thousands of class-level source files, each annotated with one of the 13 GoF design patterns labels and "Unknown" label. In the feature extraction step, we use the CodeT5+ embedding model to generate fixed-length vector representations for each source files that capture both syntactic and semantic properties of the code. This yields a \textit{Feature Corpus}, a dataset of feature vectors labeled with their corresponding pattern labels.
This corpus is split into \textit{Train set} and \textit{Test set}. The training portion is used to train a set of supervised classifiers — SVM, Logistic Regression, and MLP — constituting the \textit{DPD\textsubscript{Att}-Baseline} is explained in Chapter \ref{chp-4}. The evaluation of these classifiers on the \textit{Test set} using standard metrics such as accuracy, precision, and recall and F1-score is discussed in Chapter \ref{chp-5}.

In \textbf{Stage (2)}, to evaluate the semantic robustness, We randomly collect a subs-set of 34 java source files from the \textit{DPD\textsubscript{Att} Corpus}, sampling 2-3 files per pattern and 2 not belonging to pattern labeled "Unknown". These files are then passed through the obfuscation tool (described in detail in Chapter \ref{chp-3}), to perform systematic name obfuscation of the syntactic signals in these randomly collected source files preserving code structure and behavioral logic. Which is then collected as \textit{Obfuscated Corpus} having these source files, with each retaining its original design pattern label from \textit{DPD\textsubscript{Att} Corpus}. 
We again use the code T5+ embedding model to perform feature extraction to get a fixed-length feature vector for every obfuscated file. This is then collected in the \textit{Obfuscated Feature Corpus}. This corpus is used as \textit{Test Set} in similar fashion to stage (1) and passed through the pre-trained classifiers in \textit{DPD\textsubscript{Att}-Baseline} from stage(1) under for evaluation. The prediction results are reported in chapter \ref{chp-6}.

Finally the Comparison \& Analysis between the original and obfuscated results from stage(1) and stage (2) is presented in Chapter \ref{chp-7} respectively.

\section{Obfuscation Tool Description}\label{chp-3}

This chapter outlines systematic procedure to remove syntactic signals from the java source files while preserving the code integrity. The outline for the tool selection process and what the tool requires and how the end result would look would be described in the subsections sequentially as follows:

\subsection{Obfuscation Strategy and Objectives}\label{sub-sec3.1}
Our primary goal is to determine whether the \textit{DPD\textsubscript{Att}-Baseline} (i.e., the pre‐trained design‐pattern classifiers) relies on surface-level syntactic signals such as (class, method, and variable names) when detecting design patterns. To do so, we apply a systematic name-based obfuscation that:
\begin{itemize}
  \item Renames all identifiers in code ( such as classes, methods, variables, etc) to name-neutral placeholders.
  \item Preserves control-flow, method bodies, inheritance hierarchies, and overall program logic.
  \item Removes semantic hints embedded in print statements and comments via manual replacement.
\end{itemize}
By preserving the program logic while removing surface-level syntax, we isolate the impact of syntactic cues on classification performance.

\subsection{Obfuscation Tool selection and comparison}

To implement the obfuscation, we surveyed multiple java obfuscation tools with varying capabilities. Table 1 Summarizes the comparison:

\begin{table}[htbp]
\caption{Comparison of Obfuscation Tools}
\begin{center}
\begin{tabular}{|c|c|c|c|}
\hline
\textbf{Tools} & \textbf{License Type} & \textbf{Level} & \textbf{Mapping File} \\
\hline
Jobfuscator\cite{jobfuscator}       & Closed & Source  & No  \\
yGuard\cite{yguard}                 & Open   & Byte-code & No  \\
ProGuard\cite{proguard}             & Open   & Byte-code & Yes \\
Zelix KlassMaster\cite{zelix}       & Closed & Byte-code & Yes \\
\hline
\end{tabular}
\label{tab:obfuscation_tools}
\end{center}
\end{table}

Considering the objectives mentioned in \ref{sub-sec3.1}, we selected ProGuard due to its open-license, support for reliable bytecode-level obfuscation, and generation of detailed mapping files necessary for aligning obfuscated files with original pattern labels. However, ProGuard does not obfuscate string literals, which we addressed manually during manual code stubbing mentioned in \ref{sub-sec3.3}

\subsection{Obfuscated Corpus preparation, Manual Stubbing, and Compilation}\label{sub-sec3.3}

ProGuard tool requires compiled ".jar " file as input, necessitating successful compilation of Java source files. this proved non-trivial due to the nature of the \textit{DPD\textsubscript{Att} Corpus}: each project folder contains only those java classes that are a part of design pattern, omitting external dependencies.

While the \textit{DPD\textsubscript{Att} Corpus} was built by augmenting and refining the original DPD\textsubscript{F}- Corpus\cite{8feture-based}, the cleaned version of DPD\textsubscript{F}-Corpus which was derived from large open-source repository named GitHub Java Corpus (GJC)\cite{githubCorpus2013} still suffers from the outdated java features, unresolved imports making difficult to compile. The augmented part having additional files in \textit{DPD\textsubscript{Att} Corpus} crawled from  GitHub repositiories, does not document the specific project versions or commit hash used, making to reproduce those project structure reliably. The authors in \cite{houichime2025context} also note the inability for execution of the projects from DPD\textsubscript{F} \cite{8feture-based} and in turn same issues for DPD\textsubscript{Att}\cite{mzid2024attention} being based on DPD\textsubscript{F}.

\begin{figure}[htbp]
  \centering
  \includegraphics[width=\linewidth]{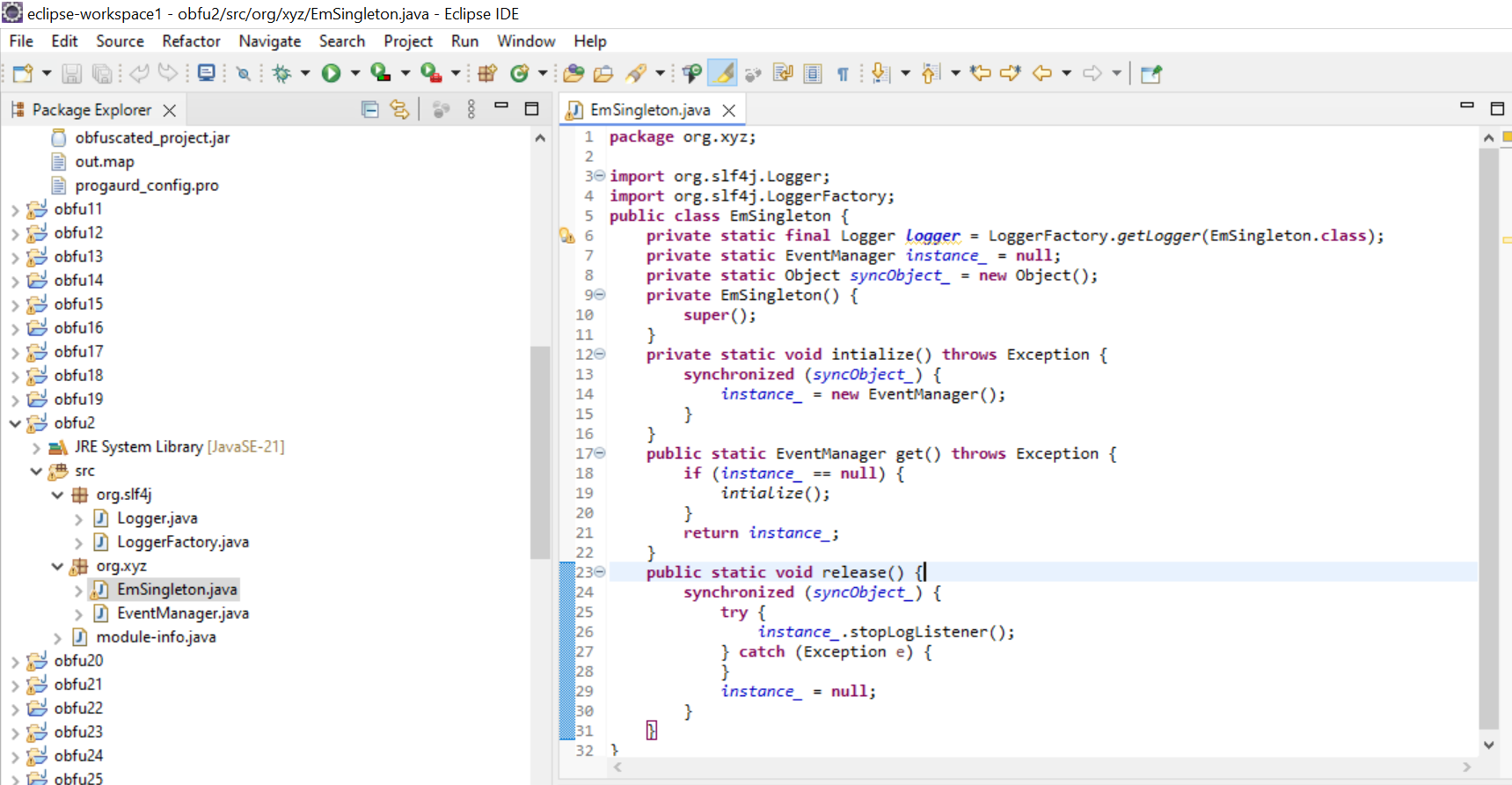}
  \caption{One of the 34 original file sample \texttt{EmSingleton.java} loaded and compiled in Eclipse IDE (prior to obfuscation).}
  \label{fig:singleton_eclipse}
\end{figure}

To overcome this, we performed manual stubbing using Eclipse IDE (as shown in Figure \ref{fig:singleton_eclipse}):
\begin{itemize}
    \item  For each selected file, we created a minimal Java Project in Eclipse IDE.
    
    \item  We manually implement and use IDE suggestions to resolve missing dependencies with stub classes or dummy methods to enable successful compilation.
    
    \item String literals and semantically revealing content (e.g., in print statements or comments) were obfuscated or removed to eliminate detection bias.
\end{itemize}

This manual process was resource-intensive and limited our scope to a representative subset of 34 Java Files (2-3 design patterns). Selection was random to avoid bias, and each file was mapped to its corresponding label using the pattern annotation  provided by the CSV file of \textit{DPD\textsubscript{Att} Corpus}.

\subsection{Obfuscation and De-compilation}\label{sub-sec3.4}

After stubbing and compilation, each project was passed through the ProGuard. This results a ".jar" and a mapping file( e.g, containing " org.xyz.EmSingleton → a.b.a") to trace the obfuscated class names. After identifying the obfuscated .class file (e.g. a.class) we de-compiled it using the FernFlower de-compiler plugin from the Visual Studio Code.

\begin{lstlisting}[style=javaStyle, caption={Obfuscated code sample labelled as Singleton from the collected subset }, label={lst:obfuscated_singleton}]
import a.a.b;

public class a {
   private static final a.a a = b.a(a.class);
   private static b.b a = null;
   private static Object a = new Object();

   private a() {
   }

   private static void b() {
      synchronized(a) {
         a = new b.b();
      }
   }

   public static b.b a() {
      if (a == null) {
         b();
      }

      return a;
   }

   public static void a() {
      synchronized(a) {
         try {
            a.a();
         } catch (Exception var2) {
         }

         a = null;
      }
   }
}
\end{lstlisting}

To illustrate, Listing \ref{lst:obfuscated_singleton} shows the de-compiled obfuscated version of "EmSingleton.java". Despite the identifier renaming, method structure, synchronization blocks, and instance management remain intact, preserving the Singleton pattern's intent.

\subsection{Final Test Set: Obfuscated corpus}
We repeated this process for mentioned above for all 34 files, resulting in an \textit{Obfuscated Corpus} with preserved pattern labels. the mapping file ensured traceability between original and obfuscated class file, enabling a valid \textit{Test Set} for classifier inference used in Chapter \ref{chp-6}.

\section{Training the Classifiers of DPDAtt-Baseline for Pattern Detection} \label{chp-4}

This chapter describes the setup and training of the DPD\textsubscript{Att}-Baseline classifiers using the feature corpus derived from the \textit{DPD\textsubscript{Att} Corpus}. It outlines the feature extraction pipeline and the supervised learning strategy employed to train the baseline models.

\subsection{Dataset and Pre-processing}\label{sub-sec4.1}

The \textit{DPD\textsubscript{Att} Corpus} consists of 212 project folders comprising 1645 Java source files. But the corpus annotations with 13 GoF design pattern labels and with "Unknown" labels was available only for 1554 files. Also the authors of {DPD\textsubscript{Att} used the same set for embedding generation pipeline, using the CodeT5+ model to generate 256-dimensional feature vectors for each file. The configuration was kept consistent with the original setup of authors. The resulting \textit{Feature Corpus}, consisting of labeled code embeddings, serves as the input for training classifiers.

\subsection{Classifier Setup}

The authors of DPD\textsubscript{Att} originally used three machine learning models for classification: Support Vector Machine (SVM), Logistic Regression, and Multi-Layer Perceptron (MLP). In our setup, we adopted the same classifiers using the scikit-learn (sklearn) library with default hyper-parameters and for MLP (max\_iter=1000) was set for convergence. These classifiers were trained in a supervised fashion using the "Train Set" which has fixed-length feature vector and pattern labels for it in the \textit{Feature Corpus}.

\subsection{Training Procedure}

The authors of DPD\textsubscript{Att} applied K-fold cross-validation (k=5), splitting the \textit{Feature Corpus} into 80\% training and 20\% testing sets across folds. Each classifier was trained and validated across five folds to ensure robustness and generalization. During each fold, the trained models were evaluated using standard classification metrics—accuracy, precision, recall, and F1 score—and subsequently saved from the last fold to be used later for evaluation in Chapter \ref{chp-6} and \ref{chp-7} respectively.

\section{Reproducing the DPDAtt-Baseline Classifiers} \label{chp-5}

This chapter presents reproducing the baseline classification performance originally reported in DPD\textsubscript{Att}\cite{mzid2024attention} As outlined in \ref{sub-sec2.2} and implemented in Chapter \ref{chp-4}, we trained three classifiers (SVM, Logistic Regression, and Multi-Layer Perceptron) on the \textit{Feature Corpus} generated from the DPD\textsubscript{Att} Java Corpus.

While the original DPD\textsubscript{Att} study reported using 300-dimensional embeddings, we were unable to replicate this setup due to the lack of reproducibility details in the configuration. The CodeT5+ encoder used in the embedding pipeline does not support an explicit configuration for a 300-dimensional output. Instead, we followed the default CodeT5+ configuration provided in \cite{mzid2024attention} , which produces 256-dimensional feature vectors. This is consistent with the behavior of the provided token configuration file by the authors.

To ensure the use of 256-dimensional vectors did not degrade performance, we trained the classifiers using the same cross-validation approach (5-fold ) described in the original paper. Table \ref{tab:classifier_results} shows a side-by-side comparison of the classification metrics obtained from the original 300 dimension embedding setup (as reported) and our 256 dimension replication.

\begin{table}[htbp]
\caption{Classifier performance comparison on 300 dimension vs 256  dimension embeddings with cross-validation k=5}
\label{tab:classifier_results}
\begin{center}
\begin{tabular}{|l|l|l|l|l|l|}
\hline
\textbf{Classifier} & \textbf{Dim} & \textbf{Accuracy} & \textbf{Precision} & \textbf{Recall} & \textbf{F1 Score}\\
\hline
SVM & 300 & 0.86 & 0.87 & 0.86 & 0.86 \\
Logistic Regression & 300 & 0.81 & 0.82 & 0.81 & 0.81 \\
MLP & 300 & 0.84 & 0.85 & 0.84 & 0.84 \\
SVM & 256 & 0.86 & 0.86 & 0.86 & 0.86 \\
Logistic Regression & 256 & 0.81 & 0.82 & 0.82 & 0.81 \\
MLP & 256 & 0.84 & 0.85 & 0.84 & 0.84 \\
\hline
\end{tabular}
\end{center}
\end{table}

\section{Testing the trained DPD\textsubscript{Att}-Baseline Under the Obfuscated Approach} \label{chp-6}

This chapter evaluates how the classifiers trained in Chapter~\ref{chp-4} perform on systematically obfuscated Java source files. The purpose of evaluation is to investigate whether these classifiers can maintain detection performance when syntactic signals are obscured, thereby testing their semantic generalization capability.

\subsection{Obfuscated Corpus and Feature Extraction} \label{sub-sec-6.1}

As described in Chapter \ref{chp-3}, we selected a representative subset of 34 Java files from the original \textit{DPD\textsubscript{Att} corpus}, with 2–3 files per GoF design pattern labels and also "Unknown" labels to files that are not a part of pattern, similar to the original corpus and call it \textit{Obfuscated Corpus}. These files were manually obfuscated to systematically rename syntactic signals (e.g., class names, method names, and variable names, etc) while preserving the structural and behavioral logic of the source code.

To create feature vectors for this new test set, we used the same embedding process described in Chapter \ref{chp-4} — the CodeT5+ encoder — to generate 256-dimensional vector representations. This resulted in the creation of \textit{Obfuscated Feature Corpus},which retains the pattern labels for the corresponding feature vectors from the \textit{Obfuscated Corpus} in similar fashion to original \textit{Feature Corpus} creation from \textit{DPD\textsubscript{Att} Corpus} mentioned in \ref{sub-sec4.1}.

\subsection{Evaluation of trained DPD\textsubscript{Att}-Baseline Under Obfuscation}

The trained classifiers in DPD\textsubscript{Att}-Baseline 
 descibed in Chapter \ref{chp-4}  were evaluated on the \textit{Obfuscated Feature Corpus}, which is used as a Test Set. Enabling a direct evaluation of the trained model’s predictions under syntactic perturbation.
The classifier metric evaluation result is shown in Table \ref{tab:Obf_classifier_results}.

\begin{table}[htbp]
\caption{Classifier evaluation metrics under obfuscated test set (34 files)} \label{tab:Obf_classifier_results}
\begin{center}
\begin{tabular}{|l|l|l|l|l|}
\hline
\textbf{Classifier} & \textbf{Accuracy} & \textbf{Precision} & \textbf{Recall} & \textbf{F1 Score}\\
\hline
SVM & 0.176 & 0.182 & 0.167 & 0.128 \\
Logistic Regression & 0.147 & 0.078 & 0.143 & 0.081\\
MLP & 0.147 & 0.190 & 0.119 & 0.127 \\
\hline
\end{tabular}
\end{center}
\end{table}
\section{ Analysis and Discussion}\label{chp-7}

Testing DPD\textsubscript{Att}-Baseline on the original \textit{Feature Corpus} confirms the high classification scores, as after running and validating with 256 embeddings-- e.g.,their best pefroming classifer SVM achieves F1=0.86 (in Chapter \ref{chp-5}; Table \ref{tab:classifier_results}). However, when exactly same models are tested on the \textit{Obfuscated Feature Corpus} (Chapter \ref{chp-6}), performance collapses across all metrics. For instance, SVM drops from F1-score=0.86 to F1-score=0.128 (chapter \ref{chp-6}; Table \ref{tab:Obf_classifier_results}). This drastic drop can be attributed to the superficial syntactic signals--such as identifier names and string literals, since control flow, inheritance and  logic were preserved during obfuscations--this answers \textbf{RQ1}.

Although the values are quite small and differ marginally in Table \ref{tab:Obf_classifier_results}. They are still reflect meaningful distinction. While SVM achieves the highest accuracy ($\approx 0.18$) and recall ($\approx 0.17$) , MLP delivers slightly better precision ($\approx 0.19$)  and a comparable F1 score (0.127) as compared to SVM's (0.128). Although the values differ marginally. This indicates that SVM retrieves more true positives but with a slight trade-off in precision, whereas MLP is more conservative yet balanced in its predictions. This may be due to its nonlinear capacity to model latent features, though the advantage remains marginal. Logistic Regression underperforms across all metrics, with poor precision and lowest F1, reflecting an overgeneralizing behavior. Collectively, these results reinforce that all models primarily exploit syntactic artifacts and possess only limited semantic generalization.

Pattern-wise confusion matrix (omitted for space) for Chapter \ref{chp-6}  reveals a consistent drift toward Prototype and Builder pattern predictions across classifiers. These patterns likely function as semantic defaults, reflecting the model's inability to disentangle structural intent or behavioral logic once name-level signals are removed. This suggests that the models fail to internalize deeper class-level semantics, leading to overgeneralized misclassification.

The 34-file obfuscated subset therefore fulfills its role as a "stress test". Because (i) it preserves code logic and structure, (ii) covers all 13 GoF patterns, and (iii) exposes a reproducible mapping file, the set offers a controlled proof-of-concept benchmark for any detectors that claims semantic awareness.
Although larger benchmarks are desirable, this subset already discriminates strongly between genuine semantic modeling and mere name-based memorization. Affirming \textbf{RQ2} as a viable benchmark for probing semantic robustness and can be reused for future work.

Finally, we note that although DPD\textsubscript{Att} compares its approach with external tools such as FeatureMaps\cite{FeatureMaps} and MARPLE-DPD\cite{zanoni2015applying} and DPD\textsubscript{F}\cite{8feture-based},  their implementation details were not provided in software artifact of DPD\textsubscript{Att} paper, preventing a like-for-like comparison on our corpus. Their omission, however, does not affect the primary finding: DPD\textsubscript{Att}’s performance depends on syntactic cues and degrades when those cues are erased.

\section{Conclusion and Future Work}\label{chp-8}

In this paper, we reproduced and extended the DPD\textsubscript{Att} design pattern detection by systematically evaluating the semantic robustness under code obfuscation. While the baseline classifiers (SVM, Logistic Regression, MLP) demonstrated strong performance on the original \textit{DPD\textsubscript{Att} Corpus}, their effectiveness dropped drastically when tested on an obfuscated subset in which naming cues were removed but structural and behavioral semantics were preserved. Thus, the degradation confirms in the current attention-based detectors like DPD\textsubscript{Att} fail to understand the code structure  and code functionality.

To support reproducible evaluation of such robustness, we introduced a manually constructed, label preserved benchmark that consists of 34 obfuscated Java files spanning 13 GoF design patterns. This small benchmark serves as effective diagnostic tool to evaluate deep semantic understanding of pattern detectors.

Future work will focus on using this benchmark to define suitable criteria for selecting detection tools that are compatible with our dataset, enabling comparative evaluation against other state-of-the-art detectors.

\begin{credits}
\subsubsection{Data Availability}
The data that support the findings of this study are available at: \url{https://github.com/manthan410/Benchmark_dpd_att}

\subsubsection{\discintname}
The authors have no competing interests to declare.

\end{credits}
%
%
%
\bibliographystyle{splncs04}
\bibliography{references}

\end{document}